\documentclass[twocolumn,preprintnumbers,amsmath,amssymb]{revtex4}
\usepackage{psfrag}
\usepackage{amssymb,amsmath,amsthm}
\usepackage[dvips]{graphicx}
\usepackage{verbatim}
\usepackage{graphicx}
\usepackage{dcolumn}
\usepackage{bm}

\begin{document}


\title{Exploring the Kibble-Zurek mechanism in a secondary bifurcation}
\author{M.A. Miranda}
\email[e-mail address:\ ]{montse@alumni.unav.es}
\affiliation{Dept. of Physics and Applied Mathematics, Universidad de Navarra. Irunlarrea s/n, E-31080 Pamplona, Spain}

\author{J. Burguete}
\email[e-mail address:\ ]{javier@fisica.unav.es}
\affiliation{Dept. of Physics and Applied Mathematics, Universidad de Navarra. Irunlarrea s/n, E-31080 Pamplona, Spain}

\author{W. Gonz\'alez-Vi\~nas}
\email[e-mail address:\ ]{wens@fisica.unav.es}
\affiliation{Dept. of Physics and Applied Mathematics, Universidad de Navarra. Irunlarrea s/n, E-31080 Pamplona, Spain}

\author{H. Mancini}
\email[e-mail address:\ ]{hmancini@fisica.unav.es}
\affiliation{Institute of Physics, University of Navarra. Irunlarrea s/n, E-31080 Pamplona, Spain}

\begin{abstract}
 We present new experimental results on the quenching dynamics of an
 extended thermo-convective system (a network array of approximately 100 convective
 oscillators) going through a secondary subcritical bifurcation. We
 characterize a dynamical phase transition through the nature of the
 domain walls (1D-fronts) that connect the basic multicellular pattern
 with the new oscillating one. Two different mechanisms of the relaxing dynamics at the threshold are characterized depending on the crossing
 rate $\mu=\left.\frac{d\varepsilon}{dt}\right|_{\varepsilon=0}$ (where
 $\varepsilon$ is the control parameter) of the quenched
 transition. From the analysis of fronts, we show that these mechanisms follow
 different correlation length scales $\xi \sim \mu^{-\sigma}$. Below a
 critical value $\mu_c$ a slow response dynamics yields a spatiotemporal
 coherent front with weak coupling between
 oscillators. Above $\mu_c$, for rapid quenches, defects are trapped at
 the front with a strong coupling between oscillators, similarly to the
 Kibble-Zurek mechanism in quenched phase transitions. These defects,
 which are pinned to the fronts, yield a strong decay of the correlation length.

\bigskip


\noindent {\small{\it Note:} A version of this article has been accepted
 for publication in the {\it Int. J. Bifurcation and Chaos}}
\smallskip
\end{abstract}

\maketitle


\section{Introduction}
 
 A dynamical synchronization transition is the closest expression of a
'real' phase transition in nature. Some recent interest on synchronization
processes~\cite{Miranda10,Mertens10,Mancini10,Arenas08,Osipov07,Abrams06,Zhou06,Boccaletti06,Shima04,Kuramoto03} is meant to
understand the interaction between oscillating units which give rise to
a collective behavior (i.e. as in a phase transition or in a Network). A key aspect of
most phase transitions is the breaking of symmetries which should be
inherited in the synchronization processes.\par

Moreover, in nature, systems that undertake symmetry breaking
transitions show a richer phenomenology. As a matter of fact, it has
been argued that in the early universe, short after the Big Bang this
kind of transitions had made possible the appearance of the present
electromagnetic field from unified fields, and the asymmetry between
matter and antimatter in the universe~\cite{Kibble64,kibble1976,Kibble80}, among other important phenomena. \citet{kibble1976} proposed that cosmological phase transitions which
underwent in the early universe could also be responsible for the large
scale structure that is observed in the present time (i.e. distribution
of galaxies). This may happen due to the existence of many causally
uncorrelated regions in a less symmetric phase after the
transitions. This will lead to a phase mismatch and thus, to the appearance of phase singularities (topological defects). Those defects determined huge energy fluctuations which could be related to the large scale of the universe, and even to the existence of topological dark matter \cite{muruyama2010}.

Later, \citet{zurek1985,zurek1996} proved that this causal mechanism has
its counterpart in condensed matter systems, and in general can be
applied to all second order breaking phase transitions. The generalized
mechanism is known as the Kibble-Zurek one. The argument is the
following: on one hand, when the control parameter is far from the
critical point, any slow change in the control parameter would be
followed adiabatically by the states of the system. On the other hand,
in the critical region the relaxation time diverges. Thus, the system
cannot follow the control parameter because otherwise the fluctuations
should propagate faster than the limiting speed in the system. Consequently the correlation length of the fluctuations gets frozen until the adiabatic dynamics is restored (after the transition is crossed). Then, fluctuations grow to form the new phase, and topological defects appear due to the aforementioned mismatch. This kind of defects, due to their topological stability, keep the correlation values until well after the transition is performed. This mechanism leads to a power law for the correlation length as a function of the rate of change of the control parameter.

Many condensed matter experiments have been performed to confirm cosmological theories in the laboratory~\cite{zurek1985,rajantie2002,rajantie2003,Dziarmaga10}. Among them, in liquid crystals~\cite{chuang1991,bowick1994,digal1999}, in superfluid helium~\cite{hendry1994,bauerle1996,ruutu1996,ruutu1998,eltsov2000,bunkov2003,eltsov2004,eltsov2010}, in superconductors and Josephson junctions~\cite{carmi1999,monaco2001,monaco2002,monaco2003,monaco2006,monaco2008,Monaco09,maniv2003}, in Bose-Einstein condensates~\cite{weiler2008,sadler2006} and in other condensed matter systems \cite{yusupov2010}. Numerical and theoretical approaches have been numerous.

The Kibble-Zurek mechanism has also been extended to non-equilibrium
primary bifurcations~\cite{casado2007} and has been the focus of
experiments in nonlinear optical systems~\cite{ducci1999,casado2007} and
in fluid convection
systems~\cite{casado2001,gonzalez2001,casado2002}. All these
bifurcations are supercritical (i.e. second order) or very weakly
subcritical. In a primary bifurcation the system goes from a homogeneous
state to a patterned one (with broken symmetries). In a secondary
bifurcation, the previous existing phase (pattern) could nonlinearly
interact with the fluctuations. Thus, the dynamics of the critical modes show features that will be relevant similarly to primary bifurcations where the broken symmetry phase has more than one critical mode~\cite{casado2001}.

On the other hand, in subcritical bifurcations we have to distinguish two cases: weakly subcritical bifurcations  where the causal mechanisms could hold for fast enough transitions, due to the existence of a slowing down of the relevant modes (although the relaxation times do not diverge) \cite{casado2001,schutzhold2008} and others where the mechanisms are completely different \cite{vachaspati2006}.

In this paper, we show first experimental results on the quenching
dynamics taking place in a convective network of oscillators driven by a quasi-1D heating
through a secondary bifurcation. The basic pattern is a stationary multicellular pattern
(ST) from which different heating ramps will send the system towards an oscillatory
pattern throughout the presence of domains of traveling waves (TW) and
mixed patterns of counter-oscillating waves over ST
(ST/ALT). Furthermore, as we increase the quench intensity the system
will cross a second bifurcation to the counter-oscillating pattern
(ALT). This experiment has been studied in a previous
experimental setup~\cite{Burguete93,Burguete03} and an improved version can be
found in~\cite{Miranda08,Miranda09}. Therefore, this experiment allows us to study the quenching
dynamics from a constrained degeneracy given by a nonhomogeneous basic
pattern (ST). For this system we have already shown that the classical
pattern formation can be understood from the point of view of
Networks~\cite{Miranda10} as the phase synchronization between
individual oscillators. In this sense, this work is very relevant
because the results on the quenched dynamics in the critical region can
be translated into the interaction and collective behavior of those
oscillators. Notwithstanding the existence of very few numerical
works~\cite{Ciszak09} pointing towards this direction.\par

Although the classical Kibble-Zurek mechanism (its limitations and
possible extensions) is still a matter of discussion
\cite{moro1999,biroli2010}, we aim at finding new clues in experiments
which could enlighten the Kibble-Zurek mechanisms from its foundations
(i.e. causality and dynamical aspects of bifurcations). Specifically, we
are going to focus for the first time on the fact that the studied
bifurcation is secondary, and also we will consider its weak (but not very weak, though measurable) subcritical character.

\section{Experimental setup}
\label{sec:setup}

 Our system consists on a rectangular cell $L_x \times L_y$ ($L_x = 470$
mm, $L_y = 60$ mm) filled with a Silicone oil of viscosity
5cSt. Convection is achieved by heating the fluid layer from
below and along a central line (in the widest direction $\hat x$). The
fluid layer lies over a plane plate with a heating rail underneath. 
Temperature on this underlying plate ($T_b$) is selected at a heating bath
($T_h$). The upper surface of the fluid
layer is opened to the atmosphere and the depth of the fluid layer $d$
is measured with a micrometric screw. Lateral cooled walls ($T_c$) and
room temperature ($T_a$) are kept at $20.0\pm0.1^{\circ}$C. The control parameter for a fixed depth is the vertical temperature
difference $\Delta T_v = T_{h} - T_a$  and
the reduced control parameter is defined as $\varepsilon=\frac{\Delta
T_v-\Delta T_{vc}}{\Delta T_v}$, where $\Delta T_{vc}=T_{hc} - T_a$ is
the critical value at the threshold. A more detailed description of the
present experimental setup can be found in~\citet{Miranda08}.
For the results reported here, we have been supplied with a different
5cSt Silicone oil whose only effect is to displace the convective thresholds.\par

During each quench, the temperature $T_b$ is recorded at the heating line and controlled at the entrance of the cell in order to check the heating bath
response. Because the underlying plate is the most external structure from
the inner core of the cell, we have obtained that $T_b\approx T_h-9.0^\circ$C.\par

The dynamics of the convective cells is recorded from the shadowgraphy images on the screen by placing an acquisition line (we use an image acquisition
board connected to a CCD camera) next to the heating line. Each
measurement consists on a spatial and temporal sampling image in grey
levels which is a spatiotemporal diagram. Spatiotemporal diagrams show a
central region of the cell of 156 mm long that is recorded during 900 seconds at a frequency of 1 s$^{-1}$.\par
Due to the high thermal inertia of this system, before recording each spatiotemporal
diagram it is necessary to achieve a permanent regime which takes
at least three hours from an initial state where the whole cell is at room temperature. 

\section{Measurement process}

\begin{figure}
\includegraphics[width=8cm]{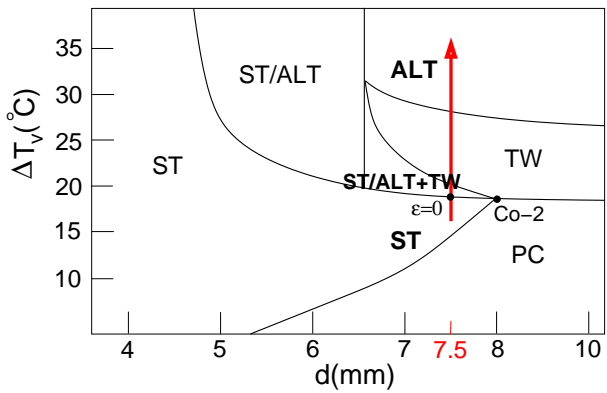}
\caption{Stability diagram. Solid lines bound regions with the same
 asymptotic dynamics. The upward arrow shows the range of quenches that
 has been explored at $d =7.5$ mm. Co-2 stands for codimension-2 point. Stationary
 patterns are: PC (primary convection which, in our system, is the
 homogeneous state) and ST (multicellular pattern). Oscillatory patterns are: ALT (resonant triad), ST/ALT (mixed pattern of irregular clusters in ALT over ST), TW (traveling waves) and ST/ALT + TW (coexistence between the mixed pattern ST/ALT and TW).}
\label{fig1}
\end{figure}

\begin{figure}
\includegraphics[width=8cm]{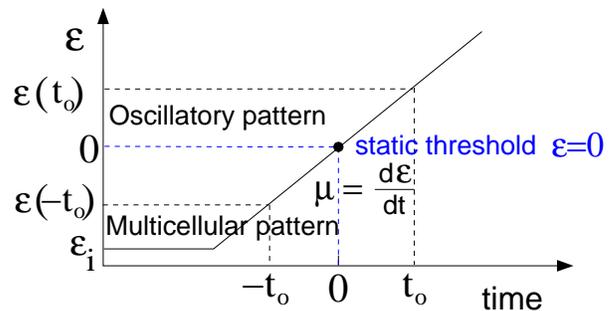}
\caption{Sketch of the control parameter ramp at the threshold of a
 secondary instability $\varepsilon=0$ $(T_b=29.5^\circ$C$)$. The slow relaxation dynamics takes place in
 the region $[-t_o,t_o]$ where the crossing rate is given by
 $\mu=\left.\frac{d\varepsilon}{dt}\right|_{\varepsilon=0}$.  The initial value of the control
 parameter for each quench is $\varepsilon_i$ $(T_{b}=27.2^\circ$C$)$ where the system exhibits the multicellular regime (ST). Above $\varepsilon=0$ the
 system bifurcates to an oscillatory pattern.}
\label{fig2}
\end{figure}

 On the stability diagram in Fig.~\ref{fig1}, we set our control
parameters at ($d= 7.5$ mm, $\Delta T_v=16.0^\circ$C), close to the codimension-2 point
from below, in order to achieve a constant velocity at the static threshold of
the secondary bifurcation $\varepsilon=0$ (see Fig.~\ref{fig2}). Thus,
the crossing rate is given by $\mu=\left.\frac{dT}{dt}\right|_{\varepsilon=0}$ . This secondary bifurcation is weakly subcritical when it is crossed quasi-statically (with a subcriticality of $\epsilon\approx-0.02$~\cite{Miranda08}).
 Under these conditions, the system bifurcates from a multicellular pattern towards the oscillatory pattern ST/ALT+TW at the first
subcritical threshold, or towards ALT at the second one depending on the
quench power. We observe from Fig.~\ref{fig1} that, beyond the
codimension-2 point ($d\ge$ 8 mm), the system bifurcates supercritically from a homogeneous state towards traveling waves.\par

The starting temperature selected at the heating bath is determined from
the closeness to the static threshold $\varepsilon=0$ of the secondary bifurcation towards
ST/ALT+TW. This fact assures the linear behavior of temperature in the
vicinity of $\varepsilon=0$. Thus, quenches start at the
stationary regime ST which is characterized by a wavelength ($\lambda_s$) that
allows us to define an array of convective oscillators, initially at rest. In the course of quenches, these convective
oscillators will become nonlocally coupled through the correlation
length $\xi$. This correlation length is measured at the
transition front.\par

Each heating ramp starts at $T_h=36.0^{\circ}$C (on the underlying plate
the corresponding temperature is $T_b=27.2\pm 0.1^{\circ}$C) and is
characterized by the final temperature in the range from
$T_h=42.0^{\circ}$C to $T_h=54.0^{\circ}$C. In order to get a better
accuracy, we have worked with steps of 1 or 2 degrees
depending on the subcritical nature of the region to be studied.\par 

The results reported here correspond to the average of four
spatiotemporal diagrams for each heating
ramp. For a given crossing rate $\mu$, we find out the
corresponding transition front. We define this 1D-front as the spatial
position where the new phase has been identified in time:
$F_\mu(x,t)=0$. From now on, it will
be expressed explicitly as $t=f_\mu(x)$, which is a univaluated function. Fronts are determined from the amplitude and phase diagrams because it is not possible to
track the oscillators paths individually along the quenched
dynamics. Trajectories described by each oscillator might become deformed
along quenched regimes invading neighboring paths in order to adjust to
a new wavenumber and frequency. This fact has determined the necessity of
analyzing the global phase-amplitude information of spatiotemporal diagrams for a constant illumination. Therefore,
the front of active oscillators $t=f_\mu(x)$ is obtained by complex
demodulation from the matrix of
amplitudes $\left|A(x,t)\right|\ge \alpha\left|A_{max}(x,t)\right|$, where $\left|A_{max}(x,t)\right|$ is the
maximum amplitude of the critical mode and $\alpha$ is a constant ($0<\alpha<1$). Critical modes
are selected with the following criteria: those with the highest amplitude in the Fourier spectrum,
and those providing the dynamics at the front.
Amplitude diagrams have been tested in the range
$\alpha=[0.2-0.37]$ in order to assure that the shape of the front is
conserved. \par
The self-correlation of the fronts is obtained as: 
\begin{equation}
\mathcal
C_\mu(\zeta)=\int_x[f_\mu(x+\zeta)-\langle f_\mu(x)\rangle][f_\mu(x)-\langle
f_\mu(x)\rangle]dx\,\,
.\label{selfcorrelation}
\end{equation}
where $\zeta$ is the spatial lag and $\langle\rangle$ defines the spatial average value.
Hence, the characteristic correlation length $\xi$ is given by the
spatial lag at $e^{-1}$ of the largest value of $\mathcal C_\mu$:
\begin{equation}
\xi=\min\left[\mathcal C_\mu ^{-1}\left\{\frac{\mathcal C_\mu(0)}{e}\right\}\right] 
\,\,
.\label{correlationlength}
\end{equation}
The values of $\mathcal C_\mu(\zeta)$ have been computed for different
parameters (amplitude thresholds and filter properties). From these
values we have obtained a similar description
of the dynamics, therefore we have introduced these deviations in the
error bars computed for each spatiotemporal diagram.\par

\begin{figure}
\includegraphics[width=8cm]{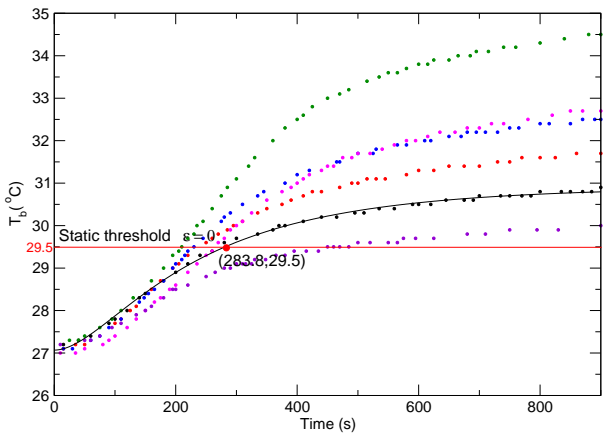}
\caption{Temperature measurements at the underlying plate of the cell
 belonging to a sequence of quenches. The static threshold of this
 secondary bifurcation $\varepsilon= 0$ takes place at
 $T_{b}=29.5^{\circ}$C. Each temperature profile has been fitted to
 the curve $T_b=a_o+a_1e^{-t/a_2}+a_1(1-e^{-t/a_2}e^{-t/a_3})$. The
 solid line corresponds to the following parameter values: $a_o= 30.91$,
 $a_1=-3.84$, $a_2=254.10$, $a_3=96.38$. The crossing rate in this particular case is $\mu=\frac{dT_{b}}{dt}(T_b=29.5^\circ$C$)=0.0061^\circ$C/s.}
\label{fig3}
\end{figure}

Temperature measurements at the underlying plate are shown in
Fig.~\ref{fig3} for a particular sequence of quenches.  We implement a
nonlinear curve fitting taking into account the heat transfer from
the water circulation in the heating bath towards the underlying plate
(see Fig.~\ref{fig3}). For each temperature ramp , the crossing rate is
obtained from the slope of each fitted curve at the static threshold: $\mu=\left.\frac{dT_{b}}{dt}\right|_{\varepsilon=0}$.\par

\section{Results and discussion}

 The initial pattern at the beginning of each quench is the stationary
one (ST), with wavelength $\lambda_s=4.75\pm0.05$ mm, which corresponds
to an array of 95 latent oscillators, although we
observe only the central region of the system (approximately 33 oscillators).
Once the quenched regime bifurcates towards an oscillatory
pattern, we obtain the corresponding front $t=f_\mu(x)$ by selecting the Fourier
oscillatory modes with the minimum frequency in order to obtain the first
unstable front from the multicellular pattern.\par

\begin{figure}
\includegraphics[width=7cm]{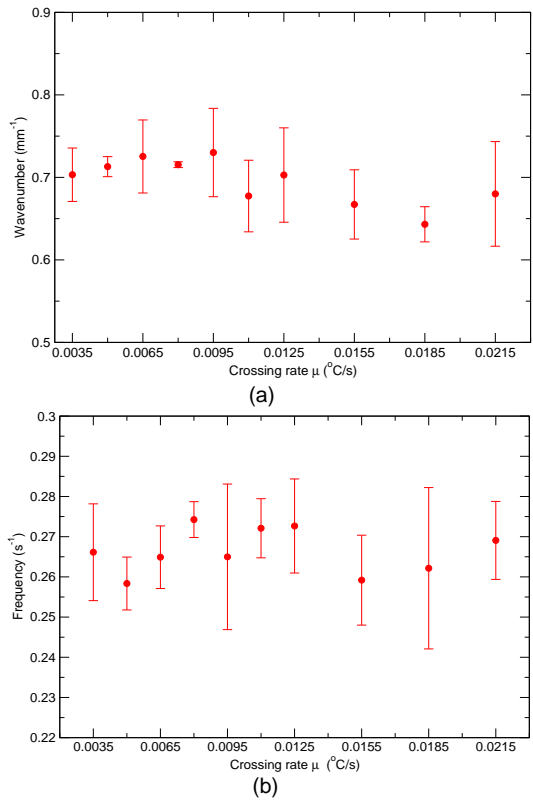}
\caption{Wavenumbers (a) and frequencies (b) of the critical modes that
 define the front at a given quench. Error bars are given by the
 dispersion of values given by the left and right oscillatory modes for
 different sequences.}
\label{fig4}
\end{figure}

In Fig.~\ref{fig4}, we show the wavenumbers and frequencies of the
critical modes. From Fig.~\ref{fig4}(a) we have
measured an average wavelength of $8.76\pm0.01$ mm for
$\mu\le0.0095^\circ$C/s, and of $9.25\pm0.02$ mm for $\mu >0.0095^\circ$C/s. The gap between
these values might be inherited from two different and consecutive subcritical
thresholds in the permanent regimes~\cite{Miranda08}. This observation will
be checked with the corresponding spatiotemporal diagrams and fronts further on.\par

From Fig.~\ref{fig4}(b) we obtain an average period of $23.60\pm0.01$ s
which remains practically constant at any crossing rate. A detailed
analysis of the spatiotemporal diagrams shows that the frequency evolves
slowly along the heating ramp as it is expected from previous results~\cite{Miranda08}. This fact is related to
the slow response of the system towards the transition point where the
modes with the lowest frequencies are the very first to guide this
quenched bifurcation. \par 

\begin{figure}
\includegraphics[width=8cm]{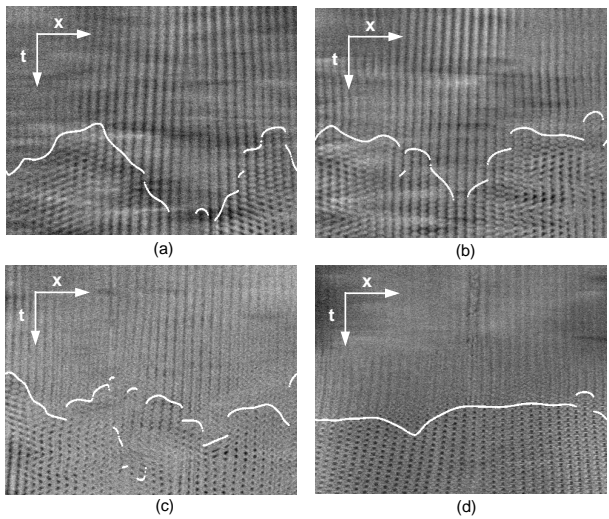}
\caption{Spatiotemporal diagrams corresponding to quenched dynamics: (a) $\mu=0.0035^\circ$C/s, (b) $\mu=0.0065^\circ$C/s, (c) $\mu=0.0095^\circ$C/s, (d) $\mu=0.0125^\circ$C/s. White lines correspond to the fronts defined by left oscillatory modes. }
\label{fig5}
\end{figure}

\begin{figure}
\includegraphics[width=8cm]{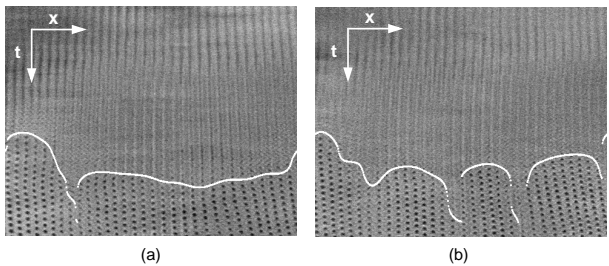}
\caption{Spatiotemporal diagrams corresponding to quenched dynamics: (a)
 $\mu=0.0155^\circ$C/s, (b) $\mu=0.0215^\circ$C/s. White lines correspond to the fronts defined by left oscillatory modes.}
\label{fig6}
\end{figure}

In Figs.~\ref{fig5},\ref{fig6} we show some spatiotemporal diagrams that
represent the quenching dynamics along two different and contiguous
subcritical thresholds. Also, the corresponding fronts $t=f_\mu(x)$ have been
overlapped.
We briefly summarize the main features of these transitions when they
are crossed quasi-statically (for a more
detailed study of the quasi-static transitions see~\citet{Miranda08}):
\begin{itemize}
\item[\bf{(I)}] In the first subcritical bifurcation, left
and right traveling waves (with $\lambda_{TW\pm}\approx 1.5\lambda_s$) are bounded
in domains which coexist with the mixed pattern ST/ALT (with $\lambda_{ALT}\approx 2 \lambda_s$).
\item[\bf{(II)}]  In the second
subcritical bifurcation, the ALT pattern is expected to coexist with the TW
pattern until, for increasing values of the control parameter, the only remaining pattern is the ALT one.
\end{itemize}

In Fig.~\ref{fig5}, we show the transition front for different crossing rates ranging from
$\mu=0.0035^\circ$C/s, with a spatio-temporal coherent front, until the
crossing rate reaches $\mu=0.0125^\circ$C/s, with a more homogeneous
front. We observe that the front becomes fragmented in between as the
quench power is increased. Thus, we may infer that, in the range shown
in this figure, the front exhibits different degrees of coherence
depending on the strength of coupling between oscillators.\par

In Fig.~\ref{fig6}, the bifurcation scenario is more robust because from
the previous quasi-homogeneous front (e.g. see Fig.~\ref{fig5}(d)) topological defects
arise at the front. These defects have survived the quenched transition and
remain pinned to the transition front.\par

From these patterns we can distinguish two different behaviors: 
\begin{itemize}
\item[\bf{(i)}] For $\mu\le 0.0095^\circ$C/s, the quenching dynamics after the
      onset of the bifurcation shows the presence of certain domains in
      the TW and ST/ALT patterns which are remnants of the subcritical
      behavior of the quasi-static transition. We should stress that the
      front describes the time at which each oscillator bifurcates
      disregarding the fluctuating behavior of the oscillators that
      belong to the mixed pattern ST/ALT (because they are susceptible of returning to the
      stationary state).
\item[\bf{(ii)}] For $\mu>0.0095^\circ$C/s, the quenching dynamics sends the system through the
      following subcritical bifurcation. Nevertheless, the new pattern has entirely
      bifurcated to the ALT pattern. Therefore, for fast quenches, during the relaxation
      time that takes the system to reach the onset of this bifurcation,
      the most unstable mode is the one that represents the ALT pattern
      (which doubles the wavelength of the multicellular pattern). 
\end{itemize}

\begin{figure}
\includegraphics[width=8cm]{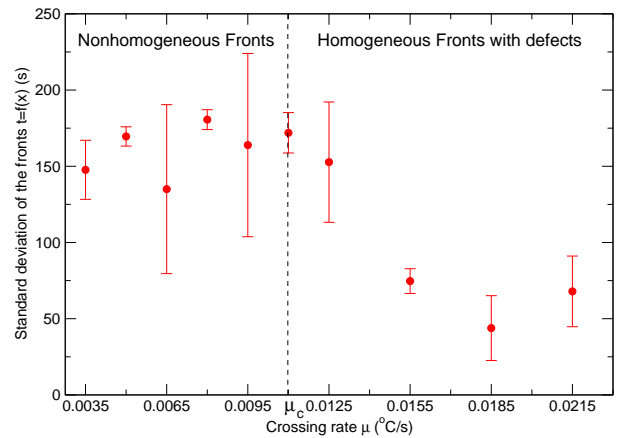}
\caption{Standard deviation of the fronts versus the crossing rate
 $\mu$. The dashed line indicates the critical value $\mu_c$.}
\label{fig7}
\end{figure}

In Fig.~\ref{fig7}, we measure for different sequences the degree of
inhomogeneity of the front given by the standard deviation of the front
$t=f_\mu(x)$. We have taken into account fronts defined from both
counter-oscillating modes. These modes define similar fronts except for the
very slow values of $\mu$ because of the presence of TW domains that are
attached to the front. We should emphasize that results from Fig.~\ref{fig7}
match the dynamical description of the fronts given above: for slow
crossing rates nonhomogeneous fronts appear, meanwhile as the crossing
rate is increased the fronts become more homogeneous and robust, this fact is related to the stronger coupling between oscillators~\cite{Miranda10}.\par

\begin{figure}
\includegraphics[width=8cm]{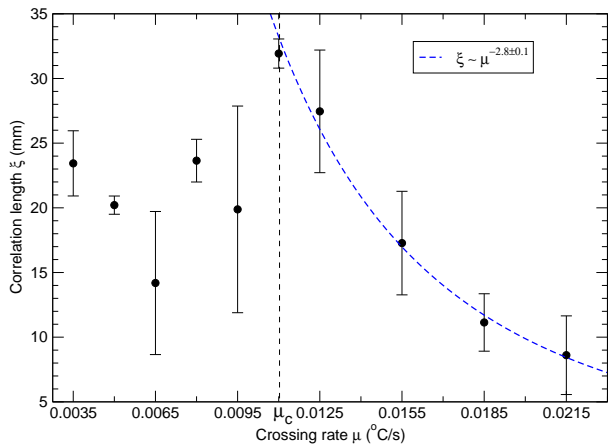}
\caption{Self-correlation of the fronts $\xi$ versus the crossing rate
 $\mu$. The vertical dashed line indicates the critical value $\mu_c$.}
\label{fig8}
\end{figure}

In Fig.~\ref{fig8}, we show the correlation length of the
fronts $\xi$ which is obtained from Eq.~(\ref{correlationlength}). The maximum
value of $\xi$ corresponds to a quasi-homogeneous front (see
Fig.~\ref{fig5}(d)) at a critical value of the crossing rate $\mu_c$. For slow quenches
($\mu<\mu_c$), there is a minimum correlation length that belongs to
the kind of fragmented fronts shown in Fig.~\ref{fig5}(b,c). In these
fronts, neighboring oscillators are beginning to show a stronger
coupling~\cite{Miranda10}. On the other hand, according to the Kibble-Zurek mechanism in phase transitions our experimental results
for fast quenches ($\mu>\mu_c$) follow a power law despite
the critical exponent is far to follow previous experimental results obtained in other systems.\par

\begin{figure}
\includegraphics[width=8cm]{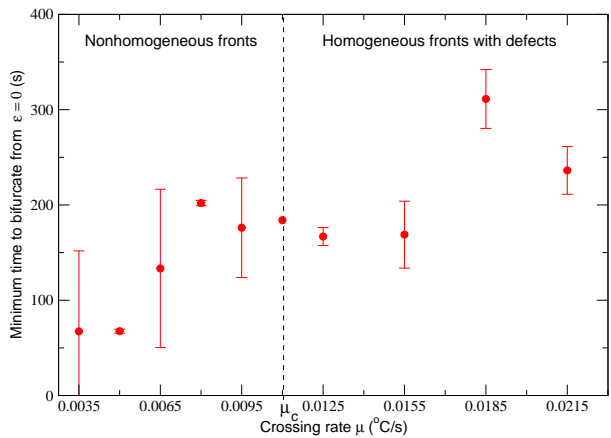}
\caption{Minimum time to bifurcate from the static threshold
 $\varepsilon=0$ versus the crossing rate $\mu$. The dashed line
 indicates the critical value $\mu_c$.}
\label{fig9}
\end{figure}

In Fig.~\ref{fig9}, the minimum time that is needed to bifurcate from
the static threshold is a measure of the relaxation time of the
convective array of oscillators. We observe that for the slowest
quenches, when the oscillators are weakly coupled, the
relaxation time is the shortest. But as the power law behavior is
reached (for $\mu>\mu_c$) the relaxation time becomes much longer as if the
quenched dynamics from the basic pattern ST to the ALT pattern had corresponded to that of a supercritical bifurcation (see Fig.~\ref{fig6}).\par

\section{Further remarks and conclusions}

 In this paper we report experimental evidence of the slow
response dynamics arising from quenched transitions in a quasi-1D
convective system. The quenched dynamics goes through a secondary and
weakly subcritical bifurcation for a depth of the fluid layer of $d=7.5$
mm. Complementary results concerning quenches through a stronger subcritical bifurcation are going to be reported elsewhere~\cite{Miranda11}.\par
 
Specifically, these results correspond to the kind of ramped symmetry breaking
bifurcations where an initially quasi-degenerated state (ST) undergoes a phase
transition where it looses some symmetries towards the oscillatory
patterns (ST/ALT+TW or ALT). This dynamics can be characterized from the shape of the
fronts where defects become trapped above a critical value of the
crossing rate $\mu_c$. Fronts and topological defects are the kind
of localized coherent structures that have survived a sudden transition in the laboratory,
 and represent an interesting analogy to the cosmological objects after the Big Bang~\cite{Kibble07}. \par 



\begin{figure}
\includegraphics[width=8cm]{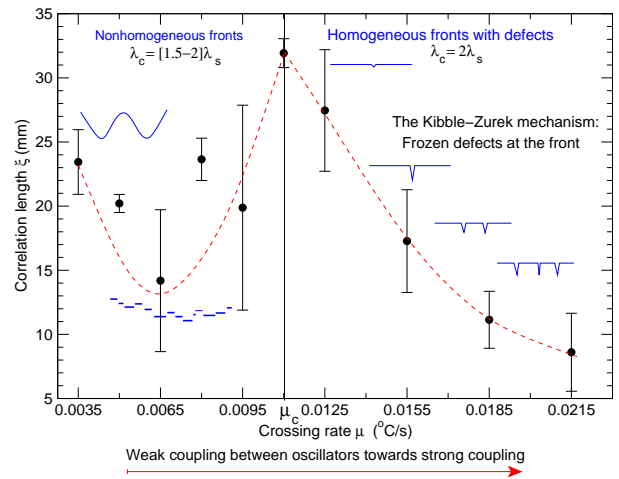}
\caption{Sketch of the quenching mechanism. Blue lines represent fronts while dashed lines
 are a guide to the eye. The vertical solid line indicates the critical
 value $\mu_c$.}
\label{fig10}
\end{figure}

According to the Kibble-Zurek mechanism, the distance between defects might
revel the order of the correlation length scale which diverges from the
correlation length $\xi$ measured from the fronts. If the positions of defects along the array are uncorrelated, one may think that the distribution of
defects along the fronts does not respond to the healing length and
perhaps to a nonlinear mechanism regarding the distant position of the
last subcritical instability from the primary bifurcation. For $\mu>\mu_c$, despite the analysis could have
been done taking as reference the next bifurcation, we choose not to;
because in fact the pattern for $\varepsilon<0$ (ST) remains unchanged
until the studied front is revealed.\par

The fronts are expected to show the degree of interaction
between the convective oscillators along the network. In
Fig.~\ref{fig10}, we summarize the most relevant aspects of this
paper. We should outline that the results on the correlation length are a
measure of the degree of interaction between oscillators. On the other
hand, fluctuations of the front that
have not been absorbed by the front during the slow relaxation time, do
represent an order parameter of the quenched dynamics. In consequence,
the shape of the front is a trace of the slow relaxation dynamics
at the critical point of a symmetry breaking bifurcation. \par

\begin{acknowledgments} 
This work has been partly
supported by the Spanish Contract No. FIS2008-01126 and by the Gobierno of Navarra (Departmento de Educaci\'on).
\end{acknowledgments}


\bibliography{bibmirandanew}

\end{document}